\documentclass[aps,twocolumn,superscriptaddress]{revtex4}
\usepackage{graphicx}
\usepackage{amsmath}
\usepackage{xcolor}
\begin{document}
\title{XUV plasmonic waveguides by near-zero index heterostructures}
\author{Luca Assogna} 
\affiliation{Department of Physical and Chemical Sciences, University of L'Aquila, Via Vetoio, 67100 L'Aquila, Italy}
\author{Carino Ferrante} 
\affiliation{CNR-SPIN, c/o Dip.to di Scienze Fisiche e Chimiche, Via Vetoio, Coppito (L'Aquila) 67100, Italy}
\author{Alessandro Ciattoni} 
\affiliation{CNR-SPIN, c/o Dip.to di Scienze Fisiche e Chimiche, Via Vetoio, Coppito (L'Aquila) 67100, Italy}
\author{Andrea Marini} 
\email{andrea.marini@univaq.it}
\affiliation{Department of Physical and Chemical Sciences, University of L'Aquila, Via Vetoio, 67100 L'Aquila, Italy}

\begin{abstract}
The lack of transmissive photonic components in the extreme ultra-violet (XUV) constitutes a challenge for micro/nano-metric confinement. 
Here, we theoretically design a novel approach to attain XUV radiation guidance based on the electromagnetic properties of 
Titanium-Aluminum-Titanium heterostructures in such a spectral domain. 
We show that, thanks to the near-zero-index properties of aluminum and titanium, XUV radiation can couple efficiently with 
plasma oscillations in such heterostructures, enabling the excitation of several distinct plasmon polariton modes. 
Our predictions, based on the semi-analytical solution of fully vectorial Maxwell's equations, indicate that the dispersion
profile of plasmon polariton modes can get efficiently modulated by the aluminum thickness, enabling nanometre confinement 
and micrometre propagation length. Moreover, we quantify the third-order nonlinearity enhancement factor, finding that it is 
resonant at the zero-index wavelength. Our results are promising for the development of future devices enabling
advanced control and manipulation of XUV radiation.
\end{abstract}

\maketitle
\section{Introduction} 
Extreme ultraviolet (XUV) radiation provides two main advantages with respect to optical radiation: (i) a shorter wavelength, enabling a more efficient lithography \cite{Wu2007} and 
a more efficient pulse compression of few-cycle pulses to achieve attosecond time-scale \cite{Krausz2001,Sekikawa2004,Sola2006,KrauszNatPhot2014,Ye2020} and (ii) a selective 
probing of specific core electrons, enabling spectroscopy with atomic selectivity \cite{Allaria2013,Balerna2019}. These properties, with the advent of pulsed sources based on high-harmonic generation 
and free-electron lasers have stimulated a large interest in physics and chemistry, providing theoretical insight of new spectroscopic schemes 
\cite{Rohringer2016,Kowalewski2017,Rohringer2019,Cavaletto2021,Nam2021}, and recent experimental implementations, e.g., nonlinear spectroscopy \cite{Cao2018}, 
coherent multidimensional spectroscopy \cite{Cao2016,Marroux2018}, soft X-ray photo-ionization \cite{Svensson2005}, 
time-resolved XUV transient absorption \cite{bencivenga2015,Drescher2016,Peng2019,Chang2020}, and time-resolved chiral order detection \cite{kerber2020}. Such emerging ultrafast techniques enable the probing of molecules and materials in 
condensed phase by photons with high energies in the XUV and are opening novel avenues to our basic understanding of ultrafast relaxation, non-equilibrium processes and 
chemical reactions \cite{Nisoli2017}. Moreover, nonlinear interaction of matter with XUV radiation enables advanced photonic applications, e.g., wave-mixing \cite{Glover2012,Foglia2018}, 
second-harmonic generation \cite{Shwartz2014,Lam2018}, saturation of absorption \cite{Mincigrucci2015} and self-induced spectral tuning \cite{Ferrante2021}. However, the inherently 
weak light-matter interaction at XUV photon energies, combined with the limited power of current XUV tabletop sources, hamper their disruptive potential for ultrafast spectroscopy 
and extreme nonlinear optics. Moreover, the high absorption in this frequency range forbids the focusing of XUV radiation with a transmissive objective, 
preventing focusing at the diffraction limit by nanometre wavelength radiation.

Radiation-matter interaction can get enhanced by plasmonic nanostructures thanks to the field enhancement produced by plasmons, collective excitations of conduction electrons 
in metals, which play a key role in surface-enhanced infrared absorption (SEIRA) \cite{Moskovits1985} and surface-enhanced Raman scattering (SERS) \cite{Moskovits2005}. Both 
techniques (SERS and SEIRA) enable chemical sensing with sensitivity reaching the single-molecule detection limit \cite{Nie1997,Kneipp1997,Rodriguez2009} and have been
adopted in several applications, e.g., pregnancy tests based on metal colloids and cancer screening \cite{Lin2014}. Furthermore, physical systems supporting slow light 
light \cite{Tsakmakidis2007,Boyd2009,Kim2012,Marini2013} naturally enhance radiation-matter interaction. In particular, near-zero-index (NZI) media can slow down light 
propagation \cite{Ciattoni2013,Daguanno2014,Newman2015,Ciattoni2016} enabling large optical nonlinearity \cite{Alam2016}, self-organization of frozen light \cite{Marini2016}, 
enhanced second and third harmonic generation \cite{Vincenti2011} and active control of tunneling \cite{Powell2009}. NZI materials can be artificially realized in the form 
of metamaterials \cite{Panoiu2006,Maas2013,Gao2013} and can also naturally exist in the form of plasmas, transparent conductors, and metals near their bulk plasma 
frequency \cite{Kinsey2019}. Thanks to their properties, NZI media are currently adopted in multiple photonic applications including sensing, guiding, vortex generation \cite{Ciattoni2017}, 
trapping and emission of visible and infrared radiation \cite{Liberal2017}.  

Here we develop a method to localize XUV radiation at the nanometre scale by the exploitation of NZI heterostructures, focusing particularly on Ti-Al-Ti (Titanium-Aluminum-Titanium) multilayers. 
Indeed, the NZI properties of Al and Ti for wavelength $\lambda < 100$ nm \cite{Kinsey2019} indicate that such heterostructures can confine radiation over a length-scale of a few nanometres. 
By solving fully vectorial Maxwell's equations semi-analytically, we obtain the transcendent dispersion relations of several transverse-electric (TE) and transverse-magnetic (TM) plasmon 
polariton modes, which we solve numerically through a Newton-Raphson algorithm. We systematically analyze the confinement properties and the attenuation length of such modes, finding that 
radiation confinement can be efficiently manipulated by the Al thickness and that TM modes offer the best effective mode length for nonlinear XUV applications.
Our results indicate that NZI index media constitute a promising platform to manipulate XUV radiation enhancing its localized interaction with matter, thus opening novel avenues for 
surface-enhanced XUV spectroscopies and extreme nonlinear optics applications. 

\begin{figure}[t]
\centering
\begin{center}
\includegraphics[width=0.5\textwidth]{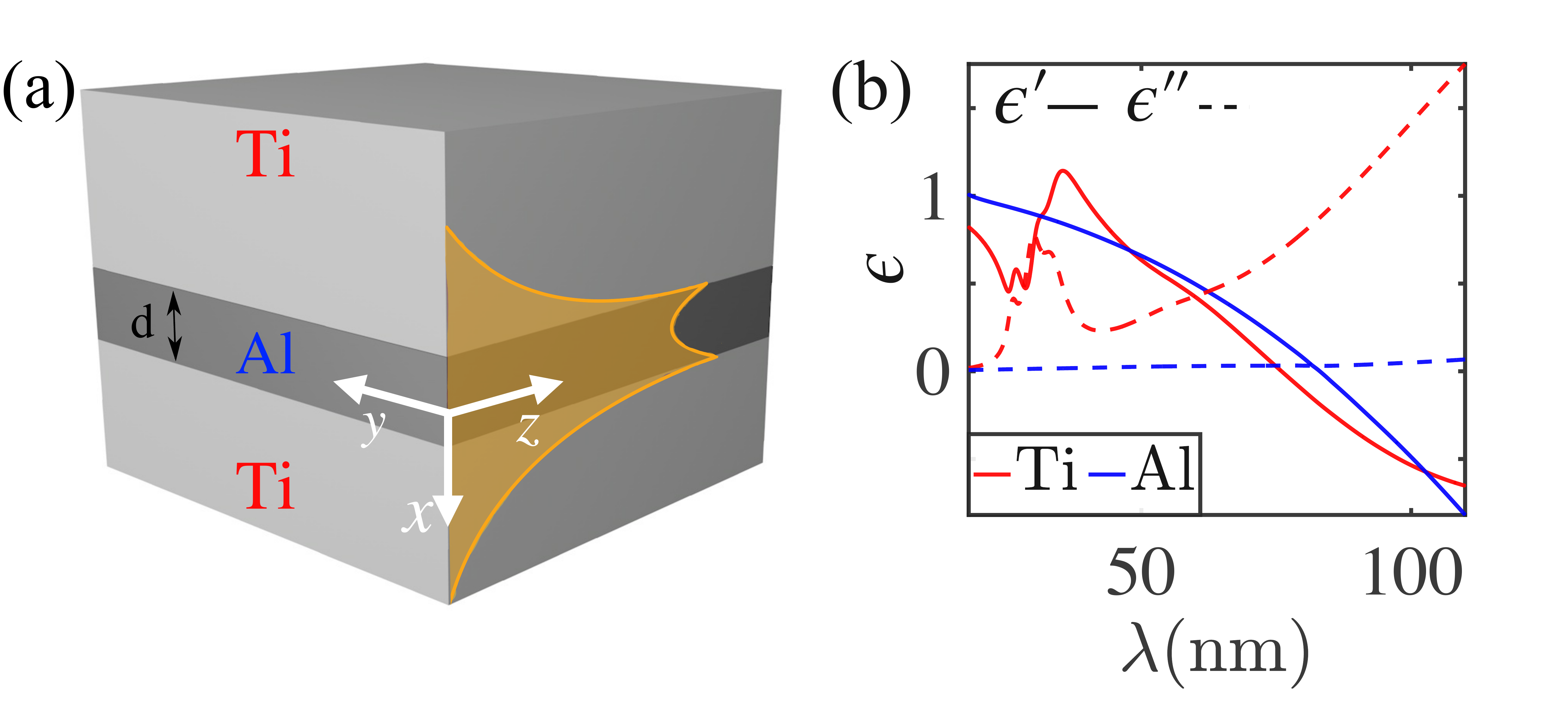}
\caption{{\bf (a)} Sketch of the considered Ti-Al-Ti heterostructure, with Al thickness $d$, supporting plasmon polariton modes propagating over the $z$-direction. 
{\bf (b)} Wavelength dependence of the complex dielectric permittivities $\epsilon_{\rm Al}$ (blue curves) and  $\epsilon_{\rm Ti}$ (red curves) \cite{Rakic1996,Werner2009}, where full/dashed curves indicate 
the real ($\epsilon'_{\rm Al,Ti}$) / imaginary ($\epsilon''_{\rm Al,Ti}$) parts of $\epsilon_{\rm Al,Ti}$.}
\label{Fig1}
\end{center}
\end{figure}

\section{Results}

We consider a Ti-Al-Ti heterostructure composed of two micron-sized Ti layers embedding a thin Al film with thickness $d\simeq 10 - 30$ nm, see Fig. \ref{Fig1}a. Owing to the large
thickness of the Ti layers compared to the XUV radiation wavelength $\lambda\simeq 10 - 100$ nm, we approximate the Ti layers as semi-infinite volumes.  
We start our analysis from the curl equations $\nabla\times {\bf B}({\bf r},t) = \mu_0 \partial_t {\bf D}({\bf r},t)$ and $\nabla\times {\bf E}({\bf r},t) = - \partial_t {\bf B}({\bf r},t)$
for the time-dependent electric and magnetic induction fields ${\bf E}({\bf r},t)$ and ${\bf B}({\bf r},t)$ at the generic position ${\bf r}$. In the linear isotropic response limit, the 
displacement vector ${\bf D}({\bf r},t)$ satisfies the constitutive relation ${\bf D}({\bf r},t) = \epsilon_0\int_{-\infty}^t \epsilon(t-t',{\bf r})  {\bf E}({\bf r},t') dt'$, where
$\epsilon(\tau,{\bf r})$ is the spatially-modulated temporal response function. The dielectric permittivity profile $\epsilon(\omega,{\bf r})$ is provided by the Fourier transform of
$\Theta(\tau)\epsilon(\tau,{\bf r})$, where $\Theta(\tau)$ is the Heaviside step function accounting for causality and $\omega$ is the radiation angular frequency. In the considered 
geometry, see Fig. \ref{Fig1}a, the dielectric permittivity profile is given by 
$\epsilon(\omega,{\bf r}) = \epsilon_{\rm Al}(\omega) \Theta_{\rm in}(x) + \epsilon_{\rm Ti}(\omega) \Theta_{\rm out}(x)$, where 
$\Theta_{\rm in}(x) = \Theta(x+d/2) - \Theta(x-d/2)$, $\Theta_{\rm out}(x) = \Theta(-x-d/2) + \Theta(x-d/2)$ and $\epsilon_{\rm Al,Ti}$ indicate
the complex dielectric permittivities of Al/Ti materials, which wavelength dependence is depicted in Fig. \ref{Fig1}b. Note that, in the XUV wavelength range $\lambda \simeq 10 - 100$ nm, 
$|\epsilon'_{\rm Al,Ti}|<1$ (where $\epsilon'$ indicates the real part of $\epsilon$) and both Al and Ti behave as NZI media \cite{Kinsey2019}.  

We seek monochromatic plasmon polariton modes with angular frequency $\omega = 2\pi c/\lambda$, where $c$ is the speed of light in vacuum, propagating over the $z$ direction and 
unbound over the $y$-direction, by taking the Ansatz ${\bf E}({\bf r},t) = {\rm Re} \left[{\bf E}_0(x,z) e^{- i\omega t} \right]$ and 
${\bf B}({\bf r},t) = {\rm Re} \left[{\bf B}_0(x,z) e^{- i\omega t} \right]$. Owing to the planar symmetry of the system, Maxwell's equations
are split into two independent sets of equations for TE and TM modes, which enable to express the complex field vectors as the superpositions
\begin{eqnarray} 
&& {\bf E}_0(x,z) = \sum_{s=\pm 1} \left[ A^{\rm (TE)}_s {\bf e}^{\rm (TE)}_s(x) e^{i \beta^{\rm (TE)}_s z} + \right. \nonumber \\
&& \left. + A^{\rm (TM)}_s {\bf e}^{\rm (TM)}_s(x)e^{i \beta^{\rm (TM)}_s z} \right],  \label{Ansatz1}\\ 
&& {\bf B}_0(x,z) = \frac{1}{c} \sum_{s=\pm 1} \left[ A^{\rm (TE)}_s {\bf b}^{\rm (TE)}_s(x)e^{i \beta^{\rm (TE)}_s z} + \right. \nonumber \\
&& \left. + A^{\rm (TM)}_s {\bf b}_s^{\rm (TM)}(x)e^{i \beta^{\rm (TM)}_s z} \right], \label{Ansatz2} 
\end{eqnarray}

\begin{figure}[t]
\centering
\begin{center}
\includegraphics[width=0.5\textwidth]{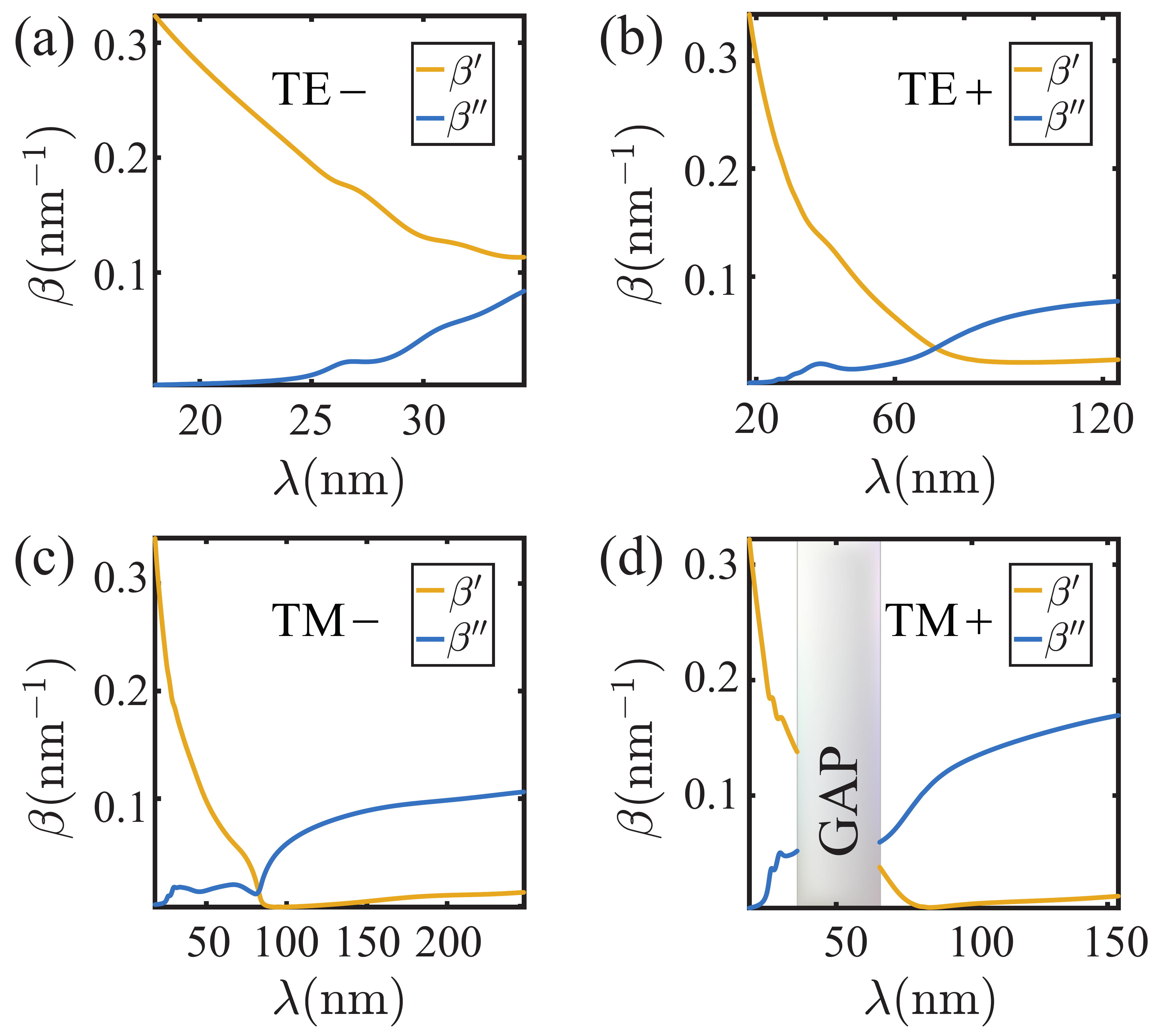}
\caption{Wavelength dependence of the propagation constant $\beta$ of lowest order modes {\bf (a)} $\beta^{\rm (TE)}_{+}(\lambda)$, {\bf (b)} $\beta^{\rm (TE)}_{-}(\lambda)$, 
{\bf (c)} $\beta^{\rm (TM)}_{+}(\lambda)$ and {\bf (d)} $\beta^{\rm (TM)}_{-}(\lambda)$ for $d = 30$ nm. In all plots the yellow/blue curves indicate the real ($\beta'$) and  
imaginary ($\beta''$) parts of the propagation constants respectively.}
\label{Fig2}
\end{center}
\end{figure}

\noindent where $A^{\rm (TE,TM)}_s$ are dimensional constants accounting for the arbitrary TE/TM mode amplitudes and ${\bf e}^{\rm (TE,TM)}_s(x)$, ${\bf b}^{\rm (TE,TM)}_s(x)$ are TE/TM dimensionless 
vector mode profiles with opposite reflection symmetry by the $y-z$ plane. Inserting the expressions above into Maxwell's equations and imposing boundary conditions for the continuity of $B_y$, 
$E_z$ and $D_x$ at the interfaces $x=\pm d/2$ together with the requirement that the field vanishes for $|x| \rightarrow +\infty$, we obtain the TM mode profiles
\begin{eqnarray} 
&& {\bf e}^{\rm (TM)}_{\pm}(x) = \left(\frac{i\beta}{q_{\rm Ti}} \hat{e}_x + \hat{e}_z \right) e^{-q_{\rm Ti}\left(x-\frac{d}{2}\right)} \Theta\left(x-\frac{d}{2}\right) + \nonumber \\
&& + \left(\mp \frac{i\beta}{q_{\rm Ti}} \hat{e}_x \pm \hat{e}_z \right) e^{q_{\rm Ti}\left(x+\frac{d}{2}\right)} \Theta\left(-x-\frac{d}{2}\right) +  \\
&& + {\cal C}_{\rm in}^{\rm (TM)} \left[ - \frac{i\beta}{q_{\rm Al}} f^{(\mp)}_{\rm Al}(x) \hat{e}_x + f^{(\pm)}_{\rm Al}(x) \hat{e}_z \right] \Theta_{\rm in}(x), \nonumber  
\end{eqnarray}
\begin{eqnarray}
&& {\bf b}^{\rm (TM)}_{\pm}(x) = \left[ \frac{i\omega\epsilon_{\rm Ti}}{cq_{\rm Ti}} e^{-q_{\rm Ti}\left(x-\frac{d}{2}\right)}\Theta\left(x-\frac{d}{2}\right) + \right. \nonumber \\
&& \left. - {\cal C}_{\rm in}^{\rm (TM)} \frac{i\omega\epsilon_{\rm Al}}{cq_{\rm Al}} f^{(\mp)}_{\rm Al}(x) \Theta_{\rm in}(x) + \right. \\
&& \left. \mp \frac{i\omega\epsilon_{\rm Ti}}{cq_{\rm Ti}} e^{q_{\rm Ti}\left(x+\frac{d}{2}\right)}\Theta\left(-x-\frac{d}{2}\right) \right] \hat{e}_y, \nonumber
\end{eqnarray}
where $\beta = \beta^{\rm (TM)}_{\pm}$ represent the propagation constants of symmetric/antisymmetric plasmon polariton TM  modes, $q_{\rm Al,Ti} = \sqrt{\beta^2 - \omega^2\epsilon_{\rm Al,Ti}/c^2}$, 
${\cal C}_{\rm in}^{\rm (TM)} = (q_{\rm Ti}\epsilon_{\rm Al} - q_{\rm Al}\epsilon_{\rm Ti})e^{-q_{\rm Al}d/2}/(q_{\rm Ti}\epsilon_{\rm Al})$, 
$f^{(\pm)}_{\rm Al}(x) = \frac{1}{2}\left(e^{q_{\rm Al}x} \pm e^{-q_{\rm Al}x}\right)$ and $\beta=\beta^{\rm (TM)}_{\pm}$ satisfies the TM dispersion relation
\begin{equation}
e^{-q_{\rm Al}d} = \pm \frac{q_{\rm Ti}\epsilon_{\rm Al} + q_{\rm Al}\epsilon_{\rm Ti}}{q_{\rm Ti}\epsilon_{\rm Al} - q_{\rm Al}\epsilon_{\rm Ti}}. \label{TMDispRel}
\end{equation}
Analogously, inserting Eqs. (\ref{Ansatz1},\ref{Ansatz2}) into Maxwell's equations and imposing boundary conditions for the continuity of $B_x$, $B_z$ and $E_y$ at the interfaces $x=\pm d/2$, 
we obtain the TE mode profiles
\begin{eqnarray} 
&& {\bf e}^{\rm (TE)}_{\pm}(x) = \left[ e^{-q_{\rm Ti}\left(x-\frac{d}{2}\right)} \Theta\left(x-\frac{d}{2}\right) + \right. \nonumber \\
&& \left. \pm e^{q_{\rm Ti}\left(x+\frac{d}{2}\right)} \Theta\left(-x-\frac{d}{2}\right) + \right. \\
&& \left. + {\cal C}_{\rm in}^{\rm (TE)} \frac{ i\omega }{ q_{\rm Al} c } f^{(\pm)}_{\rm Al}(x) \Theta_{\rm in}(x) \right] \hat{e}_y, \nonumber \\ 
&& {\bf b}^{\rm (TE)}_{\pm}(x) = - \left[ \frac{\beta c}{\omega} \hat{e}_x + \frac{q_{\rm Ti} c}{i\omega} \hat{e}_z \right] e^{-q_{\rm Ti}\left(x-\frac{d}{2}\right)} \Theta\left(x-\frac{d}{2}\right) + \nonumber \\
&& + \left[ \mp \frac{\beta c}{\omega} \hat{e}_x \pm \frac{q_{\rm Ti} c}{i\omega} \hat{e}_z \right] e^{q_{\rm Ti}\left(x+\frac{d}{2}\right)} \Theta\left(-x-\frac{d}{2}\right) + \\
&& +  {\cal C}_{\rm in}^{\rm (TE)} f^{(\mp)}_{\rm Al}(x) \Theta_{\rm in}(x)\left[ \hat{e}_z - \frac{i\beta}{q_{\rm Al}} \hat{e}_x \right] , \nonumber
\end{eqnarray}
\noindent where $\beta = \beta^{\rm (TE)}_{\pm}$ represent the propagation constants of symmetric/antisymmetric plasmon polariton TE modes, ${\cal C}_{\rm in}^{\rm (TE)} = [c(q_{\rm Al}- q_{\rm Ti})/(i\omega)] e^{-q_{\rm Al}d/2}$ and $\beta=\beta^{\rm (TE)}_{\pm}$ satisfies the 
TE dispersion relation
\begin{equation}
e^{-q_{\rm Al}d} = \pm \frac{q_{\rm Al} + q_{\rm Ti}}{ q_{\rm Al} - q_{\rm Ti} }. \label{TEDispRel}
\end{equation}
We numerically solve Eqs. (\ref{TMDispRel},\ref{TEDispRel}) for every wavelength $\lambda$ through a Newton-Raphson algorithm, obtaining the dispersion relations $\beta^{\rm (TE,TM)}_{\pm}$ 
of lowest order symmetric/antisymmetric plasmon polariton TE and TM modes, which are depicted in Fig. \ref{Fig2}a-d for $d = 30$ nm. We emphasize that, although we find also higher-order modes 
in the considered Ti-Al-Ti heterostructure, they are highly lossy and we do not report them here. Furthermore we note that, while TE modes exist over all the considered XUV wavelength range, 
symmetric ($s=+1$) TM modes are cut-off within a specific spectral gap around $\lambda \simeq 50$ nm, see Fig. \ref{Fig2}d, owing to resonant absorption of Ti, see Fig. \ref{Fig1}b. 

\begin{figure}[t]
\centering
\begin{center}
\includegraphics[width=0.5\textwidth]{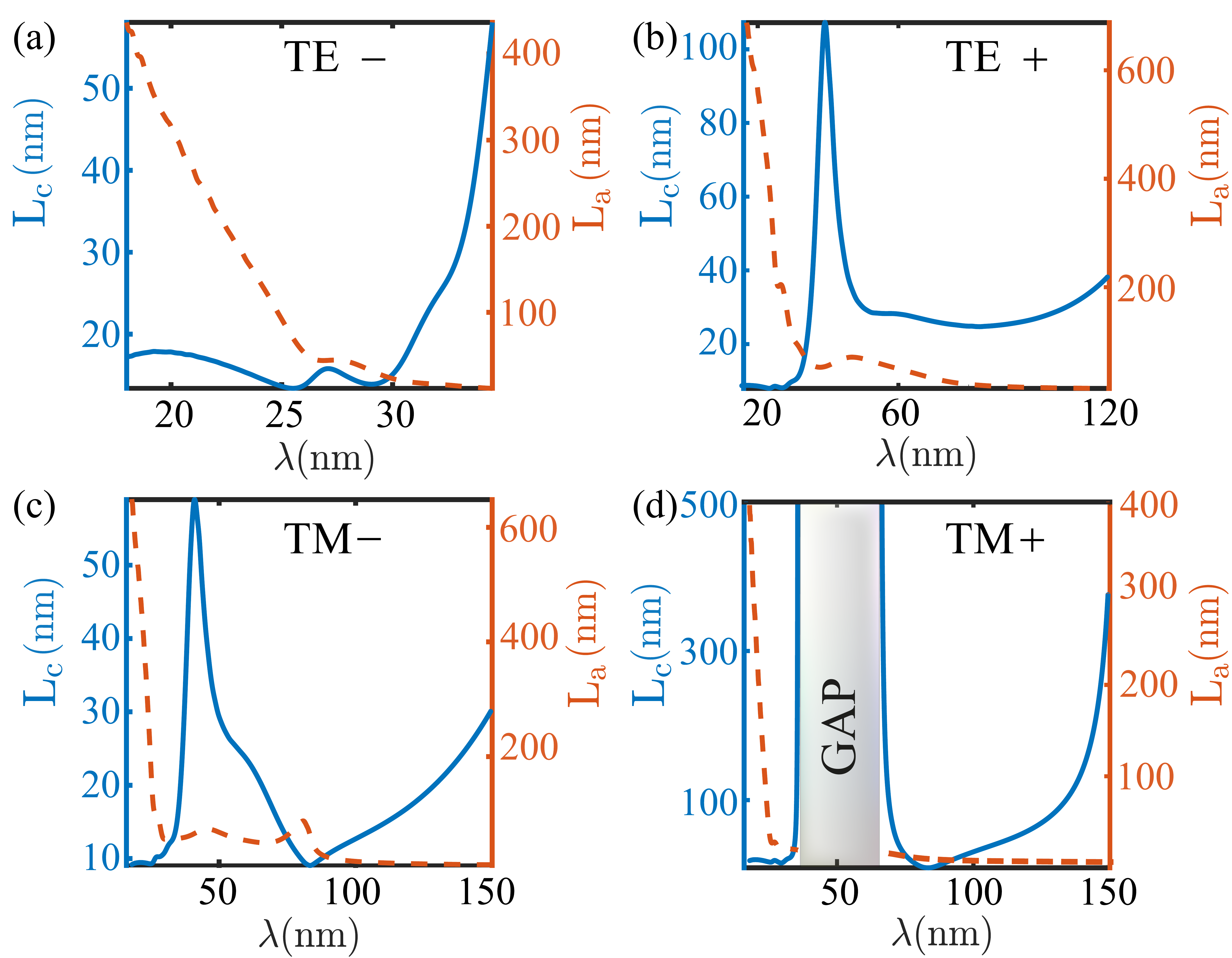}
\caption{Wavelength dependence of the attenuation $L_{\rm a}(\lambda)$ (dashed orange lines) and confinement $L_{\rm c}(\lambda)$ (full blue lines) lengths of {\bf (a,b)} TE, {\bf (c,d)} TM, {\bf (a,c)} antisymmetric ($s=-1$) and ({\bf b,d}) symmetric ($s=+1$) modes for $d = 30$ nm.}
\label{Fig3}
\end{center}
\end{figure}

\section{Discussion}

In order to quantify the propagation features of the considered plasmonic modes, we define the confinement length 
\begin{equation}
L_{\rm c} = \sqrt{ \left. \int_{-\infty}^{+\infty}dx x^2 \left|{\bf E}\right|^2 \right/ \int_{-\infty}^{+\infty}dx \left|{\bf E}\right|^2 }, 
\end{equation}
and the attenuation length $L_{\rm a} = 1/\beta''$ (where $\beta''$ indicates the imaginary part of the propagation constant $\beta$), which wavelength dependencies for TE/TM symmetric ($s=+1$)/antisymmetric ($s=-1$) modes are depicted in Fig. \ref{Fig3}.

\begin{figure}[t]
\centering
\begin{center}
\includegraphics[width=0.5\textwidth]{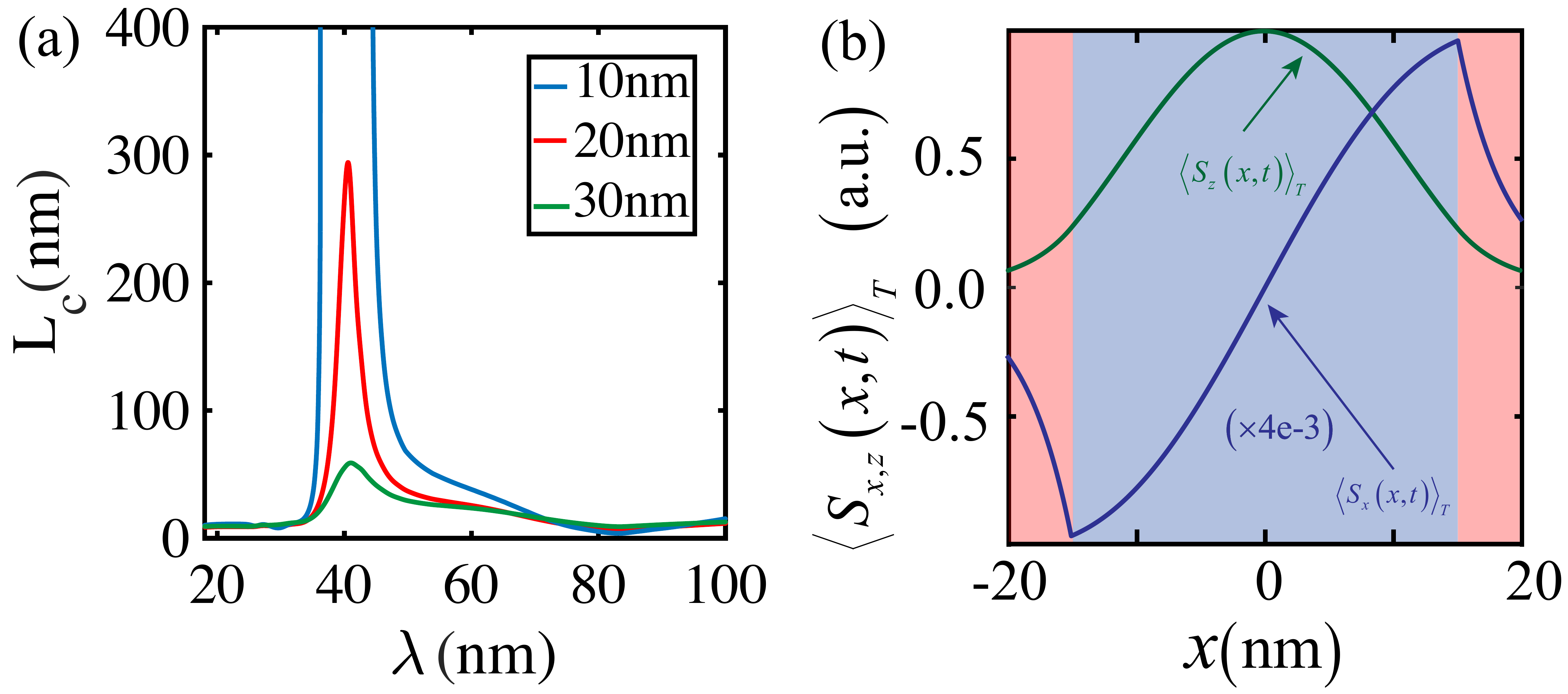}
\caption{{\bf (a)} Wavelength dependence of the confinement length $L_{\rm c}$ of TM antisymmetric ($s=-1$) modes for several distinct Al thicknesses $d = 10,20,30$ nm. {\bf (b)} Spatial dependence of the Poynting vector components $\langle S_{x,z}(x,z=0,t)\rangle_T$ of the TE+ mode averaged over the single-cycle time duration $T$ for $d = 30$ nm and $\lambda = 18$ nm.}
\label{Fig4}
\end{center}
\end{figure}

The wavelength dependence of $L_{\rm a}(\lambda)$ and $L_{\rm c}(\lambda)$ for TE/TM antisymmetric ($s=-1$) and symmetric ($s=+1$) modes is illustrated in Fig. \ref{Fig3} for $d = 30$ nm. We note that, for wavelength $17 < \lambda < 25$ nm, the attenuation length becomes very large $L_{\rm a} \simeq 500$ nm ($L_{\rm a} > 20 \lambda$) owing to the reduced absorption of Ti and Al in such
wavelength range. Conversely, for $\lambda < 17$ nm, the photon energy overcomes the ionization threshold of Al, thus leading to enhanced absorption. For every distinct mode, radiation 
confinement is maximised at peculiar wavelengths where the confinement length reaches its minimum. In turn, this peculiarity enables the tailoring of the confinement features by selective mode excitation.
Moreover, the symmetric TM+ mode ($s=+1$) enables efficient spectral control of radiation confinement thanks to the sudden increase of $L_{\rm c}$ nearby the cut-off wavelengths, 
see Fig. \ref{Fig3}d, where the confinement length diverges. Furthermore, adjusting the thickness $d$ allows for the tuning of the spectral bandgap width, thus providing a useful knob to control
radiation confinement. For instance, Fig. \ref{Fig4}a shows that the antisymmetric TM- mode ($s=-1$) is subject to a spectral bandgap opening around $\lambda \simeq 40$ nm for $d < 15$ nm, with a strong  wavelength dependence of $L_{\rm c}$ for distinct Al thicknesses $d$. Note that the confinement length at $\lambda \simeq 40$ nm can get tuned efficiently within $50$ nm $<L_{\rm c}<400$ nm by tailoring the Al thickness within $10$ nm $<d<30$ nm.

\begin{figure}[t]
\centering
\begin{center}
\includegraphics[width=0.5\textwidth]{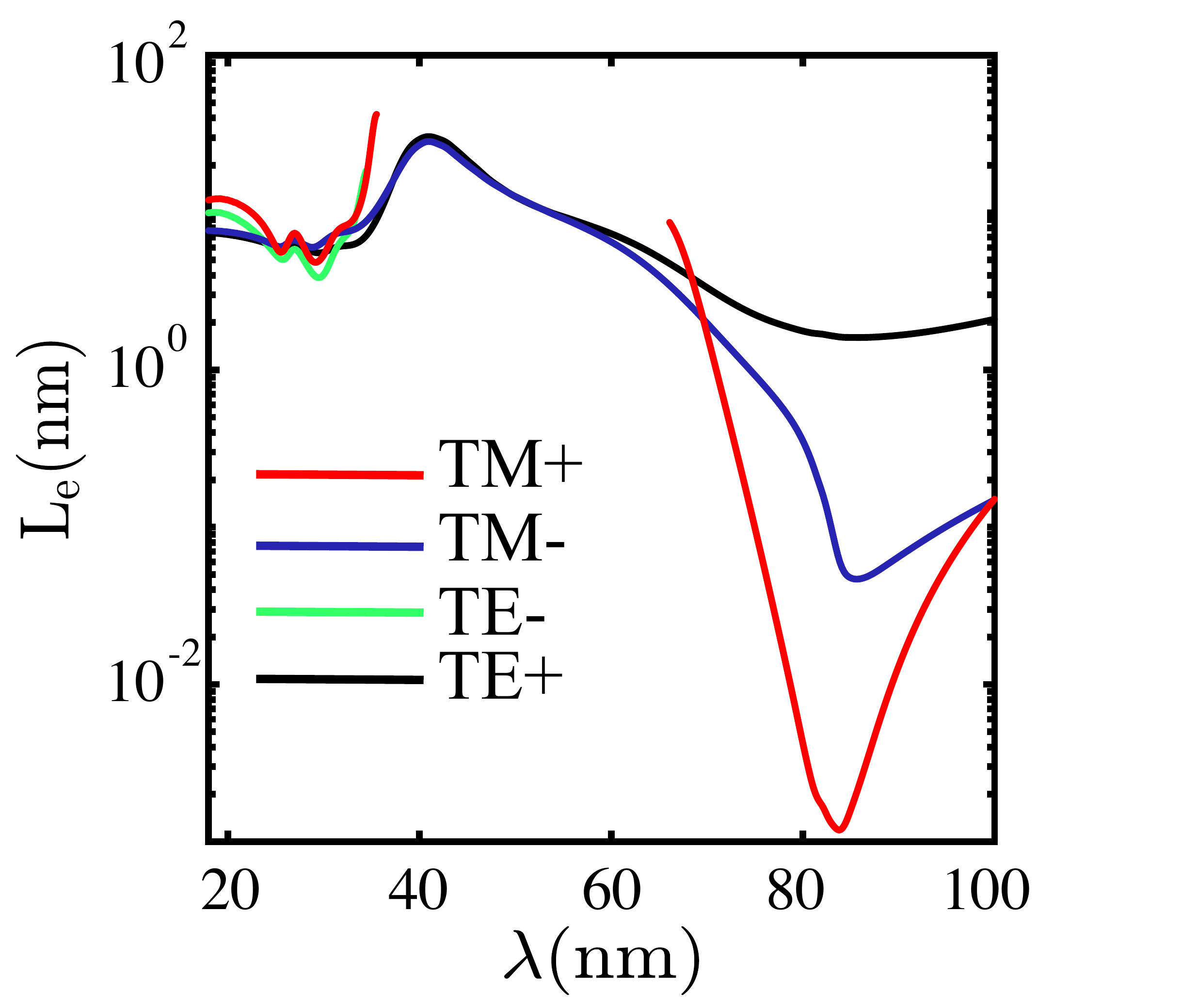}
\caption{Wavelength dependence of the effective-mode length $L_{\rm e}$ of symmetric ($s=+1$) and antisymmetric ($s=-1$) TM/TE modes for $d = 30$ nm.}
\label{Fig5}
\end{center}
\end{figure}

We emphasize that XUV radiation guidance is attained only thanks to the NZI properties of Al and Ti in the considered spectral range, which enable the excitation of plasmon polariton modes. 
In turn, electromagnetic propagation in the NZI regime implies peculiar radiation dynamics where spatially-distributed inhomogeneous absorption plays an important role, which becomes 
particularly evident in the 
power flux, accounted by the time-averaged Poynting vector ${\bf S}({\bf r,t}) = \langle {\bf E}({\bf r,t})\times{\bf H}({\bf r,t}) \rangle_T$, see Fig. \ref{Fig4}b, where $T = 2\pi/\omega$ 
is the single-cycle duration.      
Note that, while the longitudinal component of the Poynting vector ($S_z({\bf r,t}) = \langle [{\bf E}({\bf r,t})\times{\bf H}({\bf r,t})]\cdot\hat{e}_z \rangle_T$) is always positive owing to forward
plasmon polariton propagation, the transverse component ($S_x({\bf r,t}) = \langle [{\bf E}({\bf r,t})\times{\bf H}({\bf r,t})]\cdot\hat{e}_x \rangle_T$) is antisymmetric over the transverse position $x$, see Fig. \ref{Fig4}b. Such behavior ensues from the large absorption of Ti (see $\epsilon''(\lambda)$ in Fig. \ref{Fig1}b), which deflects the radiation towards the waveguide sides, and is observed
also for the other modes considered in our analysis.   
In order to quantify the enhancement factor of third-order nonlinear effects we introduce the effective mode length \cite{Marini2011}
\begin{equation}
L_{\rm e} = \frac{\left\{ \int_{-\infty}^{+\infty} {\rm Re}[ {\bf e}_s^{\rm TM,TE} \times {\bf b}_s^{\rm TM,TE *} ]\cdot\hat{e}_z dx \right\}^2}{\int_{-\infty}^{+\infty} \left|{\bf e}_s^{\rm TM,TE}\right|^4 dx},
\end{equation}
which wavelength dependence is depicted in Fig. \ref{Fig5}. Note that $L_{\rm e}$ reaches sub-nm values for symmetric/antisymmetric TM modes at $\lambda \simeq 80$ nm, where the real part of the
Al dielectric constant vanishes, see Fig. \ref{Fig1}b. Indeed, in such regime, NZI media efficiently enhance third-order nonlinearity \cite{Alam2016} owing to slow-light propagation dynamics 
enabling larger radiation-matter interaction time \cite{Ciattoni2013}.

\section{Conclusions}

In conclusion, we find that NZI hetero-structures constitute a promising platform to attain spatial confinement of XUV radiation at the nanometre scale. Moreover, the complex dispersion 
of plasmon polariton modes allows for wavelength filtering, opening new possibilities for a fine tuning of radiation localization by Al thickness manipulation. In particular, such a tunability 
can get strongly enhanced by operation nearby photonic band-gaps, where the confinement length is sharply sensitive to the Al thickness and the radiation wavelength. Furthermore, our results
indicate that mode selection plays a relevant role in the tailoring of confinement and third-order nonlinear properties owing to polarization-dependent dispersion. In particular, we find that 
the effective mode length reaches sub-nm scale for TM modes at the wavelength where the real part of the dielectric permittivity of Al vanishes, producing a giant enhancement of third-order
nonlinearity. We emphasize that XUV guidance does not rely only on Al/Ti heterostructures, but can be attained also by other NZI media, e.g., Tungsten and Vanadium. Our results
are promising for the design of novel XUV spectroscopic devices exploiting radiation confinement to enhance radiation-matter interaction and achieve nanometre spatial resolution, thus opening 
novel avenues for single-molecule detection, extreme nonlinear optics, and scanning probe microscopy with nanometre sensitivity.


\begin{thebibliography}{10}
\newcommand{\enquote}[1]{``#1''}

\bibitem{Wu2007} B. Wu and A. Kumar, {\it J. Vac. Sci. Technol. B} {\bf 25}, 1743 (2007). 

\bibitem{Krausz2001} M. Hentschel, R. Kienberger, C. Spielmann, G. A. Reider, N. Milosevic, T. Brabec, P. Corkum, U. Heinzmann, M. Drescher, and F. Krausz, {\it Nature} {\bf 414}, 509 - 513 (2001).

\bibitem{Sekikawa2004} T. Sekikawa, A. Kosuge, T. Kanai, and S. Watanabe, {\it Nature} {\bf 432}, 605 - 608 (2004).

\bibitem{Sola2006} I. J. Sola et al., {\it Nat. Phys.} {\bf 2}, 319 - 322 (2006).

\bibitem{KrauszNatPhot2014} F. Krausz \& M. I. Stockman, {\it Nat. Photon.} {\bf 8}, 205 - 213 (2014).

\bibitem{Ye2020} P. Ye et al., {\it J. Phys. B: At. Mol. Opt. Phys.} {\bf 53}, 154004 (2020).

\bibitem{Allaria2013} E. Allaria et al., {\it Nat. Commun.} {\bf 4}, 2476 (2013).

\bibitem{Balerna2019}  A. Balerna et  al., {\it Condens. Matter} {\bf 4}, 30 (2019).

\bibitem{Rohringer2016} V. Kimberg and N. Rohringer, {\it Struct. Dyn.} {\bf 3}, 034101 (2016).

\bibitem{Kowalewski2017} M. Kowalewski, B. P. Fingerhut, K. E. Dorfman, K. Bennett, and S. Mukamel, {\it Chem. Rev.} {\bf 117}, 12165 - 12226 (2017).

\bibitem{Rohringer2019} N. Rohringer, {\it Phil. Trans. R. Soc. A} {\bf 377}, 20170471 (2019).

\bibitem{Cavaletto2021} S. M. Cavaletto, D. Keefer, and S. Mukamel, {\it Phys. Rev. X} {\bf 11}, 011029 (2021).

\bibitem{Nam2021} Y. Nam, D. Keefer, A. Nenov, I. Conti, F. Aleotti, F. Segatta, J. Yong Lee, M. Garavelli, and S. Mukamel, {\it J. Phys. Chem. Lett.} {\bf 12}, 12300 - 12309 (2021).

\bibitem{Cao2018} W. Cao, E. R. Warrick, A. Fidler, S. R. Leone, and D. M. Neumark, {\it Phys. Rev. A} {\bf 97}, 023401 (2018).

\bibitem{Cao2016} W. Cao, E. R. Warrick, A. Fidler, D. M. Neumark, and S. R. Leone, {\it Phys. Rev. A} {\bf 94}, 053846 (2016).

\bibitem{Marroux2018} H. J. B. Marroux, A. P. Fidler, D. M. Neumark, and S. R. Leone, {\it Sci. Adv.} {\bf 4}, eaau3783 (2018).

\bibitem{Svensson2005} S. Svensson, {\it J. Phys. B: At. Mol. Opt. Phys.} {\bf 38}, S821 - S838 (2005).

\bibitem{bencivenga2015} F. Bencivenga et al., {\it Nature} {\bf 520}, 205–208 (2015).

\bibitem{Drescher2016} L. Drescher, M. C. E. Galbraith, G. Reitsma, J. Dura, N. Zhavoronkov, S. Patchkovskii, M. J. J. Vrakking, and J. Mikosch, {\it J. Chem. Phys.} {\bf 145}, 011101 (2016).

\bibitem{Peng2019} P. Peng, C. Marceau, M. Herv\'e, P. B. Corkum, A. Y. Naumov, and D. M. Villeneuve, {\it Nat. Commun.} {\bf 10}, 5269 (2019).

\bibitem{Chang2020} K. F. Chang, M. Reduzzi, H. Wang, S. M. Poullain, Y. Kobayashi, L. Barreau, D. Prendergast, D. M. Neumark, and S. R. Leone, {\it Nat. Commun.} {\bf 11}, 4042 (2020).

\bibitem{kerber2020} N. Kerber et al., {\it Nat. Comm.} {\bf 11}, 6304 (2020) .

\bibitem{Nisoli2017} M. Nisoli, P. Decleva, F. Calegari, A. Palacios, and F. Mart\'in, {\it Chem. Rev.} {\bf 117}, 10760 - 10825 (2017).

\bibitem{Glover2012} T. Glover et al., {\it Nature} {\bf 488}, 603 - 608 (2012).

\bibitem{Foglia2018} L. Foglia et al., {\it Phys. Rev. Lett.} {\bf 120}, 263901 (2018).

\bibitem{Shwartz2014} D. S. Shwartz et al., {\it Phys. Rev. Lett.} {\bf 112}, 163901 (2014).

\bibitem{Lam2018} R. K. Lam et al., {\it Phys. Rev. Lett.} {\bf 120}, 023901 (2018).

\bibitem{Mincigrucci2015} R. Mincigrucci et al., {\it Phys. Rev. E} {\bf 92}, 011101 (2015).

\bibitem{Ferrante2021} C. Ferrante et al., {\it Light Sci. Appl.} {\bf 10}, 92 (2021).

\bibitem{Moskovits1985} M. Moskovits, {\it Rev. Mod. Phys.} {\bf 57}, 783 - 826 (1985).

\bibitem{Moskovits2005} M. Moskovits, {\it J. Raman Spectrosc.} {\bf 36}, 485 - 496 (2005).

\bibitem{Nie1997} S. Nie and S. R. Emory, {\it Science} {\bf 275}, 1102 - 1106 (1997).

\bibitem{Kneipp1997} K. Kneipp et al., {\it Phys. Rev. Lett.} {\bf 78}, 1667 - 1670 (1997).

\bibitem{Rodriguez2009} L. Rodr\'iguez-Lorenzo et al., {\it J. Am. Chem. Soc.} {\bf 131}, 4616 - 4618 (2009).

\bibitem{Lin2014} D. Lin et al., {\it J. Biomed. Nanotechnol.} {\bf 10}, 478 - 484 (2014).

\bibitem{Marini2013} A. Marini and F. Biancalana, {\it Phys. Rev. Lett.} {\bf 110}, 243901 (2013).

\bibitem{Tsakmakidis2007} K. L. Tsakmakidis, A. D. Boardman and O. Hess, {\it Nature} {\bf 450}, 397 - 401 (2007).

\bibitem{Boyd2009} R. W. Boyd, {\it J. of Mod. Opt.} {\bf 56}, 1908 - 1915 (2009).

\bibitem{Kim2012} K.-H. Kim, A. Husakou and J. Herrmann, {\it Opt. Express} {\bf 20}, 25790 - 25797 (2012).

\bibitem{Ciattoni2013}  A. Ciattoni, A. Marini, C. Rizza, M. Scalora and F. Biancalana, {\it Phys. Rev. A} {\bf 87}, 053853 (2013).

\bibitem{Daguanno2014} G. D'Aguanno et al., {\it Phys. Rev. B} {\bf 90}, 054202 (2014).

\bibitem{Newman2015} W. D. Newman et al., {\it ACS Photon.} {\bf 2}, 2 - 7 (2015).

\bibitem{Ciattoni2016} A. Ciattoni, C. Rizza, A. Marini, A. Di Falco, D. Faccio, and M. Scalora, {\it Laser Photonics Rev.} {\bf 10}, 517 - 525 (2016).

\bibitem{Alam2016} M. Z. Alam, I. De Leon and R. W. Boyd, {\it Science} {\bf 352}, 795 - 797 (2016).

\bibitem{Marini2016} A. Marini and F. J. Garc\'ia de Abajo, {\it Sci. Rep.} {\bf 6}, 20088 (2016).

\bibitem{Vincenti2011} M. A. Vincenti, D. de Ceglia, A. Ciattoni, and M. Scalora, {\it Phys. Rev. A} {\bf 84}, 063826 (2011).

\bibitem{Powell2009} D. A. Powell et al., {\it Phys. Rev. B} {\bf 79}, 245135 (2009).

\bibitem{Panoiu2006} N. C. Panoiu, R. M. Osgood Jr., S. Zhang, and S. R. J. Brueck, {\it J. Opt. Soc. Am. B} {\bf 23}, 506 - 513 (2006).

\bibitem{Maas2013} R. Maas, J. Parsons, N. Engheta, and A. Polman, {\it Nat. Photon.} {\bf 7}, 907 - 912 (2013).

\bibitem{Gao2013} J. Gao, L. Sun, H. Deng, C. J. Mathai, S. Gangopadhyay, and X. Yang, {\it Appl. Phys. Lett.} {\bf 103}, 051111 (2013).

\bibitem{Kinsey2019} N. Kinsey, C. DeVault, A. Boltasseva, and V. M. Shalaev, {\it Nat. Rev. Mater.} {\bf 4}, 742 - 760 (2019).

\bibitem{Ciattoni2017} A. Ciattoni, A. Marini, C. Rizza, {\it Phys. Rev. Lett.} {\bf 118}, 104301 (2017).

\bibitem{Liberal2017} I. Liberal and N. Engheta, {\it Nat. Photon.} {\bf 11}, 149 - 158 (2017).

\bibitem{Rakic1996} A. D. Rakic, {\it Appl. Opt.} {\bf 34}, 4755 - 4767 (1995).

\bibitem{Werner2009} W. S. M. Werner, K. Glantschnig, C. Ambrosch-Draxl, {\it J. Phys Chem Ref. Data} {\bf 38}, 1013 - 1092 (2009).

\bibitem{Marini2011} A. Marini, R. Hartley, A. V. Gorbach, and D. V. Skryabin, {\it Phys. Rev. A} {\bf 84}, 063839 (2011).

\end{thebibliography}
\end{document}